# Many-body theory of dilute gas condensates – derivation of a field-modified Gross-Pitaevskii equation from multipolar QED

L. G. Boussiakou, C. R. Bennett, and M. Babiker

Department of Physics, University of York, Heslington, York YO10 5DD, England.

**Abstract**

The Hamiltonian of a moving atom in electromagnetic fields includes velocity-dependent terms. We show that the leading velocity dependence emerges systematically in the non-relativistic limit from a scheme firmly based on the relativistic invariance of the energy-momentum stress tensor of the coupled matter-fields system. We then extend the Hamiltonian to the many-body situation suitable for describing a Bose-Einstein condensate (BEC). From first principles, we use the equation of motion for the condensate wavefunction to obtain an extended version of the Gross-Pitaevskii (GP) equation and an equation for the internal states of the atoms. It is shown that laser fields modify the GP equation by inclusion of convective terms involving a Röntgen interaction plus a term coupling the centre of mass momentum to the Poynting vector. We also obtain the modified Maxwell equations for the electromagnetic fields coupled to the BEC involving the average velocity of the atoms.

## 1. Introduction

In a recent letter, Leonhardt and Piwnicki [1] examined magnetic effects associated with the motion of a Bose-Einstein condensate, regarded as a quantum dielectric. Earlier, it has been pointed out that a neutral atom bearing an electric dipole moment moving in an external magnetic field can accumulate a quantum phase [2-6] and a moving dipole may, under suitable conditions, exhibit a detectable Aharanov-Bohm phase shift [7]. These and other effects associated with atomic motion have received considerable attention recently, most notably in the areas of atom optics [8,9] and laser cooling and trapping [10]. The theoretical framework describing relevant phenomena involving moving atoms relies on a seemingly straightforward extension of non-relativistic quantum optics by including the translational motion of the atomic centre of mass. Clearly, it is this feature of the theory which ensures that the exchange of energy and linear momentum is properly taken into account in any interaction between the laser light and the atomic translational and internal degrees of freedom.

The need to incorporate the centre of mass motion in quantum optics theory had necessitated a re-appraisal of the corresponding quantum electro-dynamical theory where investigations sought to elucidate how the division of the motion into centre of mass and internal motions is affected by the presence of the interaction with electromagnetic fields [11-14]. One of the main outcomes of these investigations was the emphasis on the role of the Röntgen interaction energy term which couples the electric dipole moment to an effective electric field involving the centre of mass velocity and the magnetic field. Although the existence of the Röntgen interaction in quantum electrodynamics was pointed out earlier, its practical implications and its significance for the consistency of any treatment incorporating translational effects in the coupling of electric dipoles to electromagnetic fields have only relatively recently been explored [1-7,15-17].

Our first goal here is to incorporate the motion of the centre of mass in a systematic manner, aiming to derive a multipolar Hamiltonian which exhibits its velocity dependence. This is safely done using a relativistic starting point. Having obtained the velocity dependent multipolar Hamiltonian, we take the non-relativistic limit and incorporate the many body effects appropriate for describing a Bose-Einstein condensate (BEC). We must also add the S-wave scattering potential to the Hamiltonian to take into account the weak hard sphere interactions which occur in a dilute BEC [18,19]. In a similar manner to other works [18], we find the equation of motion for the condensate wavefunction assuming that the majority of the bosons are in their ground state. This leads to a modified Gross-Pitaevskii equation which now incorporates velocity dependent terms in the coupling of the atoms to the electromagnetic fields. As stated elsewhere [20], we must be careful how the polarization terms are included and divide the Coulomb interaction terms into intra- and inter-atom interactions. After taking the dipole approximation [21] and assuming that the electron position vector can be averaged in the interaction term, we may uncouple the non-linear Schrödinger equation into an equation for the gross motion of the atoms and another for the internal degrees of freedom.

In [1] it was shown how the electromagnetic fields can be modified in the presence of the BEC by including the average velocity of the atoms. Here we show how Maxwell's equations can be obtained as Heisenberg equations using the Hamiltonian we have derived and show that the sign of the velocity in [1] is different an observation which can be justified further by inspection of the Röntgen interaction. We conclude by breifly discussing the results obtained.

## 2. Transformation of the Hamiltonian

We start by writing the fully relativistic Lagrangian of the system in a frame of reference moving with the atom including the electromagnetic field and its interaction with the Dirac field for a bound electron. A contribution due to the massive nucleus is included, with the nucleus taken to coincide with the centre of mass, providing the source of the binding Coulomb field and is assumed to be moving relative to the laboratory observer. From the Lagrangian in the centre of mass frame it is straightforward to derive the stress energy-momentum tensor $T^{\mu\nu}$, with the $T^{00}$ tensor component identified as the Hamiltonian of the system in the centre of mass frame. Under a Lorentz transformation from the centre of mass frame to the laboratory frame the stress energy-momentum tensor transforms like a second rank four-tensor

$$T^{\mu\nu} = \Lambda^{\mu}_{\alpha} \Lambda^{\nu}_{\beta} T^{\alpha\beta}, \qquad (1)$$

where $\Lambda^{\mu}_{\alpha}$ is the Lorentz transformation matrix with an arbitrary relative velocity vector **V**. The required Hamiltonian in the laboratory frame emerges as $T^{00}$ and can be written as a sum of a contribution due to the electromagnetic field, a contribution due to the Dirac field together with the centre of mass term and, finally, a contribution describing interactions. The velocity dependence is manifest in various terms of this Hamiltonian.

The system Hamiltonian in the laboratory frame is next subjected to two sets of transformations. First, an exact mutipolar gauge transformation is applied and, second, the non-relativistic limit is taken. In the process of dividing the resulting Hamiltonian into unperturbed and interaction terms we retain terms to first order in the velocity. This velocity is identified with the momentum operator of the centre of mass.

The result can be written as

$$H = H_f + H_a + H_{int}. \tag{2}$$

The first term is the usual electromagnetic Hamiltonian plus a term which represents the kinetic energy arising from the transport of the field Poynting vector at velocity **V**,

$$H_f = \frac{\varepsilon_0}{2}\int\left[E^{\perp^2}(\mathbf{r}) + c^2 B(\mathbf{r})\right]d^3r - \frac{\varepsilon_0}{M}\int \mathbf{P}\cdot\left[\mathbf{E}^{\perp}(\mathbf{r})\times\mathbf{B}(\mathbf{r})\right]d^3r. \tag{3}$$

Here $\mathbf{E}^{\perp}$ is the transverse part of the electric field and **P** is the centre of mass momentum of the atom which is given by $\mathbf{V}/M$. The second term contains the gross kinetic energy of the atom, the kinetic energy of the electron and the polarization term containing the interaction between them in the Power-Zienau-Woolley representation [21],

$$H_a = \frac{P^2}{2M} + \frac{p^2}{2m} + \frac{1}{2\varepsilon_0}\int p^2(\mathbf{r},\mathbf{R},\mathbf{l})d^3r, \tag{4}$$

$$p(\mathbf{r},\mathbf{R},\mathbf{l}) = e\int_0^1 \mathbf{l}\delta(\mathbf{r} - \mathbf{R} - \lambda\mathbf{l})d\lambda, \tag{5}$$

**p** is the momentum of the electron which is at position $\mathbf{R}+\mathbf{l}$ with the centre of mass of the atom (assumed to be at the position of the proton) at **R**. The last term in Eq. (2) contains the interaction between the atom and the electromagnetic fields,

$$H_{int} = -\int p^{\perp}(\mathbf{r},\mathbf{R},\mathbf{l})\mathbf{E}^{\perp}(\mathbf{r})d^3r + \frac{1}{M}\int \mathbf{P}\cdot\left[p^{\perp}(\mathbf{r},\mathbf{R},\mathbf{l})\times\mathbf{B}(\mathbf{r})\right]d^3r. \tag{6}$$

In the above we have disregarded all magnetic multipolar interactions as well as spin interactions but the full electric multipoles as in Eq. (5) are retained at this stage.

## 3. The condensate wavefunction

We now transform the Hamiltonian in Eq. (2) to the case where there are many atoms, suitable for the case of a Bose-Einstein condensate (BEC). The boson field operator describing the internal as well as the gross motion of the ensemble can be written as a sum over the product states as follows,

$$\hat{\Psi}^{\dagger}(\mathbf{R},\mathbf{l}) = \sum_{n1,n2} \psi_{n1}^{*}(\mathbf{R})\chi_{n2}^{*}(\mathbf{l})\hat{a}_{n1,n2}^{\dagger},$$
$$\hat{\Psi}(\mathbf{R},\mathbf{l}) = \sum_{n1,n2} \psi_{n1}(\mathbf{R})\chi_{n2}(\mathbf{l})\hat{a}_{n1,n2},$$
(7)

where $\psi_{n1}(\mathbf{r})$ are the state functions associated with the centre of mass of the atom, $\chi_{n2}(\mathbf{l})$ are associated with the internal states of the atom and the operators $\hat{a}_{n1,n2}^{\dagger}$ and $\hat{a}_{n1,n2}$ are the usual boson operators satisfying the usual commutation relations. Thus, instead of Eq. (2) we have the many-body field theoretic Hamiltonian

$$H = \int\int \left\{ \begin{array}{l} \hat{\Psi}^{\dagger}(\mathbf{R},\mathbf{l})\left[\dfrac{P^2}{2M} + \dfrac{p^2}{2m}\right]\hat{\Psi}(\mathbf{R},\mathbf{l}) + \\ \dfrac{1}{M}\int\left[\hat{\Psi}^{\dagger}(\mathbf{R},\mathbf{l})\mathbf{P}\hat{\Psi}(\mathbf{R},\mathbf{l})\right]\left[\mathcal{P}^{\perp}(\mathbf{r})\times\mathbf{B}(\mathbf{r}) - \varepsilon_0 \mathbf{E}^{\perp}(\mathbf{r})\times\mathbf{B}(\mathbf{r})\right]d^3r + \\ \dfrac{1}{2}\int\int \hat{\Psi}^{\dagger}(\mathbf{R},\mathbf{l})\hat{\Psi}^{\dagger}(\mathbf{R},\mathbf{l})V_0(\mathbf{R}-\mathbf{R'})\hat{\Psi}(\mathbf{R'},\mathbf{l'})\hat{\Psi}(\mathbf{R'},\mathbf{l'})d^3R'd^3l' + \\ \hat{\Psi}^{\dagger}(\mathbf{R},\mathbf{l})V_{trap}(\mathbf{R})\hat{\Psi}(\mathbf{R'},\mathbf{l'}) \end{array} \right\} d^3Rd^3l +$$

$$\dfrac{1}{2\varepsilon_0}\int \mathcal{P}^2(\mathbf{r})d^3r - \int \mathcal{P}(\mathbf{r})\mathbf{E}^{\perp}(\mathbf{r})d^3r + \dfrac{\varepsilon_0}{2}\int\left[E^{\perp^2}(\mathbf{r}) + c^2 B(\mathbf{r})\right]d^3r,$$
(8)

where we have introduced the polarization field operator $\mathcal{P}(\mathbf{r})$ in the form

$$\mathcal{P}(\mathbf{r}) = \int\int e\int_0^1 \mathbf{l}\hat{\Psi}^{\dagger}(\mathbf{R},\mathbf{l})\delta(\mathbf{r}-\mathbf{R}-\lambda\mathbf{l})\hat{\Psi}(\mathbf{R},\mathbf{l})d\lambda d^3Rd^3l.$$
(9)

We have added two terms to this Hamiltonian. The term involving $V_0(\mathbf{R}-\mathbf{R}')$ is the scattering term due to hard-sphere interactions [18,19]. This is usually taken to be in the S-wave approximation [18,19] which is

$$V_0(\mathbf{R}-\mathbf{R}') = \frac{2\pi\hbar^2 a}{M}\delta(\mathbf{R}-\mathbf{R}') = V_0\delta(\mathbf{R}-\mathbf{R}'), \qquad (10)$$

where $a$ is the scattering length. The second term we have added is due to the trapping potential $V_{trap}(\mathbf{R})$ which is usually taken to be parabolic.

We now derive the equation of motion for $\hat{\Psi}(\mathbf{R},\mathbf{l})$ using the Heisenberg picture $-i\hbar\dot{\hat{\Psi}} = [H,\hat{\Psi}]$. Dropping all velocity dependent terms and terms containing the polarization, we have, using $\dot{\hat{\Psi}} = -iE/\hbar\,\hat{\Psi}$ where $E$ is the total energy of the state,

$$\left[\frac{P^2}{2M} + \frac{p^2}{2m} + V_{trap}(\mathbf{R}) + V_0\hat{\Psi}^\dagger(\mathbf{R},\mathbf{l})\hat{\Psi}(\mathbf{R},\mathbf{l})\right]\hat{\Psi}(\mathbf{R},\mathbf{l}) = E\hat{\Psi}(\mathbf{R},\mathbf{l}), \qquad (11)$$

which, apart from the implicit dependence on the internal states $\chi(\mathbf{l})$ and the kinetic energy term for the electrons, is reminiscent of the Gross-Pitaevskii (GP) equation prior to assuming that all particles are in the ground state.

We already see that Eq. (11) requires at least the Coulomb interaction between the proton and electron of the same atom to account for the correct internal states. This Coulomb term comes from the longitudinal part of the squared polarization term, $\frac{1}{2\varepsilon_0}\int \mathcal{P}^{\|2}(\mathbf{r})d^3r$, in the Hamiltonian. The transverse part, $\frac{1}{2\varepsilon_0}\int \mathcal{P}^{\perp 2}(\mathbf{r})d^3r$, amounts to a self-energy of the atom which we can ignore. We must, however, add the Coulomb interaction between the proton and electron to the full Hamiltonian, Eq. (8), and subtract it. After some lengthy algebra, we obtain the new form for the Schröndinger equation, which on restricting to the electron dipole approximation where $\mathcal{P}(\mathbf{r}) = e \int\int \mathbf{l} \hat{\Psi}^\dagger(\mathbf{R},\mathbf{l}) \delta(\mathbf{r}-\mathbf{R}) \hat{\Psi}(\mathbf{R},\mathbf{l}) d^3R d^3l$, becomes

$$\left[ \begin{array}{l} \dfrac{P^2}{2M} + \dfrac{p^2}{2m} - \dfrac{e^2}{4\pi\varepsilon_0 l} + \dfrac{e}{\varepsilon_0} \int d\mathbf{l}' \left[ 2\hat{\Psi}^\dagger(\mathbf{R},\mathbf{l}')\hat{\Psi}(\mathbf{R},\mathbf{l}') - \hat{\Psi}(\mathbf{R},\mathbf{l}')\hat{\Psi}^\dagger(\mathbf{R},\mathbf{l}') \right] d^3l' + \\ V_{trap}(\mathbf{R}) + V_0 \hat{\Psi}^\dagger(\mathbf{R},\mathbf{l})\hat{\Psi}(\mathbf{R},\mathbf{l}) - \mathbf{d}.\mathbf{E}^\perp(\mathbf{R}) + \left[ \mathbf{d}\times\mathbf{B}(\mathbf{R}) - \varepsilon_0 \mathbf{E}^\perp(\mathbf{R})\times\mathbf{B}(\mathbf{R}) \right] \dfrac{\mathbf{P}}{M} \end{array} \right] \hat{\Psi}(\mathbf{R},\mathbf{l}) = E\hat{\Psi}(\mathbf{R},\mathbf{l}).$$

(12)

where we have put $\mathbf{d} = e\mathbf{l}$. Next we make the assumption that the number of atoms in the condensate is very large and that they all occupy the ground state without any fluctuations. Thus, the ground state operators reduce to c-numbers, $\hat{a}_{0,0} \approx \hat{a}^\dagger_{0,0} \approx \sqrt{N}$, with all other operators $\hat{a}_{n_1,n_2}, \hat{a}^\dagger_{n_1,n_2}$ $(n_1, n_2 > 0)$ giving zero. Even so, in Eq. (12) the internal and gross motions are still coupled. To uncouple the two motions, we assume that $\mathbf{l}$ can be averaged in the interaction term and the dipole moment. We then get the total energy as $E = \mu + E_{\text{int}}$, where $\mu$ is the chemical potential of the atoms appearing in a modified Gross-Pitaevskii equation,

$$\left[ \begin{array}{l} \dfrac{P^2}{2M} + N|\psi_0(\mathbf{R})|^2 \left[ \dfrac{e}{\varepsilon_0} \int d\mathbf{l}' |\chi_0(\mathbf{l}')|^2 d^3l' + V_0 \right] + V_{trap}(\mathbf{R}) - \\ \mathbf{d}.\mathbf{E}^\perp(\mathbf{R}) + \left[ \mathbf{d}\times\mathbf{B}(\mathbf{R}) - \varepsilon_0 \mathbf{E}^\perp(\mathbf{R})\times\mathbf{B}(\mathbf{R}) \right] \dfrac{\mathbf{P}}{M} \end{array} \right] \psi_0(\mathbf{R}) = \mu \psi_0(\mathbf{R}), \quad (13)$$

and $E_{\text{int}}$ is the internal energy of the atoms which appears in the equation describing the internal degrees of freedom,

$$\left[\frac{p^2}{2m} - \frac{e^2}{4\pi\varepsilon_0 l}\right]\chi_0(\mathbf{l}) = E_{\text{int}}\chi_0(\mathbf{l}). \tag{14}$$

If we assume that there is a maximum correlation between the atoms, Eq. (13) becomes

$$\left[\begin{array}{l}\dfrac{P^2}{2M} + N|\psi_0(\mathbf{R})|^2\left(\dfrac{d^2}{\varepsilon_0} + V_0\right) + V_{trap}(\mathbf{R}) - \\ \mathbf{d}.\mathbf{E}^\perp(\mathbf{R}) + [\mathbf{d}\times\mathbf{B}(\mathbf{R}) - \varepsilon_0\mathbf{E}^\perp(\mathbf{R})\times\mathbf{B}(\mathbf{R})]\dfrac{\mathbf{P}}{M}\end{array}\right]\psi_0(\mathbf{R}) = \mu\psi_0(\mathbf{R}). \tag{15}$$

Thus, the dipole-dipole interaction term and the last three terms in the bracket are new terms in the non-linear Schrödinger equation. The dipole-dipole term, shown here to arise systematically from the theory., has previously been added phenomenologically [20,22]. However, to the authors' knowledge the **P**-dependent interaction terms involving **E** and **B** fields have not been derived rigorously before.

## 4. The modified Maxwell equations

We now turn our attention to the electromagnetic fields and the Maxwell equations associated with them. To this end we use the Heisenberg picture with the operators **B** and $\mathbf{D} = \varepsilon_0 \mathbf{E}^\perp$ and the commutation relation [23],

$$\left[\varepsilon_0 E_i^\perp(\mathbf{r}), B_j(\mathbf{r}')\right] = -i\hbar\varepsilon_{ijk}\frac{\partial}{\partial r_k}\delta(\mathbf{r}-\mathbf{r}'), \tag{16}$$

where $\varepsilon_{ijk}$ is 1 if *ijk* is cyclic, 0 if any of the indices are the same or –1 otherwise. We use this commutation relation because the fields are those in the laboratory frame in the transformed Hamiltonian and they have not coupled to the polarization. Thus we obtain the following Maxwell equations

$$\nabla \times \mathbf{H} = \frac{\partial \mathbf{D}}{\partial t}, \quad \nabla \times \mathbf{E} = -\frac{\partial \mathbf{B}}{\partial t}, \tag{17}$$

where the fields **H** and **D** contain polarization and velocity dependence as follows,

$$\mathbf{H}(\mathbf{r}) = \mu_0 \mathbf{B}(\mathbf{r}) - \varepsilon_0 N \mathbf{v} \times \mathbf{E}^\perp(\mathbf{r}), \tag{18}$$

$$\mathbf{D}(\mathbf{r}) = \varepsilon_0 \mathbf{E}(\mathbf{r}) + \mathcal{P}^\perp(\mathbf{r}) - \varepsilon_0 N \mathbf{v} \times \mathbf{B}(\mathbf{r}), \tag{19}$$

with the average velocity of the atoms given by

$$\mathbf{v} = \frac{1}{M} \int \psi_0^*(\mathbf{R}) \mathbf{P} \psi_0(\mathbf{R}) d^3 \mathbf{R}. \tag{20}$$

In the electric dipole approximation, and assuming that all the atoms occupy the ground state, the polarization, Eq. (9), is given by

$$\mathcal{P}(\mathbf{r}) = N \mathbf{d} |\psi_0(\mathbf{r})|^2. \tag{21}$$

Eqs. (17-21) are similar to the Maxwell equations derived in [1] except for the sign of the average velocity vector. It should be noted that the starting point in [1] os a model Lagrangian and the equations derived from it have the Röntgen interaction with a different sign from that published elsewhere [11,13-15]. Here, the equations arise from a theory based on a well-know Lagrangian and the associated Hamiltonian which we then systematically reduced to yield a theory for a BEC in the presence of external fields.

## 5. Conclusions

We have rigorously derived the non-linear Schrödinger equation governing the ground state wavefunction of a BEC including the dipole-dipole interactions and the internal degrees of freedom of the atoms. The interaction of the atoms with external electromagnetic fields is derived here including a BEC for the first time, appearing in direct and velocity dependent terms. We have also obtained the associated Maxwell equations for the electromagnetic fields and shown that these equations include terms dependent on the average velocity of the atoms and their polarization. The influence of the new interaction terms on the properties of a BEC in external fields has yet to be investigated, but it appears that they become significant when vortices are present [1]. This will not be pursued any further here.

**Acknowledgements**

L. G. Boussiakou would like to thank the University of York for the studentship and C. R. Bennett would like to thank EPSRC for funding.